\documentclass[twocolumn,showpacs,fleqn,nobibnotes]{revtex4}
\usepackage{amsmath,amssymb}
\usepackage{epsfig}
\usepackage{graphicx}

\newcommand{\pg}{p_{\perp}^{\gamma}}
\newcommand{\pq}{p_{\perp}^{Q}}

\begin{document}
\title{\bf Photon plus heavy quark production in high energy collisions within the target rest frame formalism}
\author{M.A. Betemps $^{a}$ and M.V.T. Machado $^{b}$}

\affiliation{
$^a$ Instituto Federal de Educa\c{c}\~ao, Ci\^encia e Tecnologia Sul-Rio-Grandense. Campus Pelotas -  Visconde da Gra\c{c}a. Av. Ildefonso Sim\~oes Lopes, 2791. CEP 96060-290, Pelotas, RS, Brazil\\
$^b$  High Energy Physics Phenomenology Group, GFPAE  IF-UFRGS.
Caixa Postal 15051, CEP 91501-970, Porto Alegre, RS, Brazil
}

\begin{abstract}
We apply the target rest frame formalism to photon $+$ heavy quark production cross section in hadronic collisions at high energies. We investigate the dependence of the production
cross section on the photon and quark rapidities and transverse momenta. It is shown that the photon transverse momentum spectrum is a sensitive probe
of color dipole scattering amplitude. The theoretical results are compared to Tevatron  measurements of the differential $\gamma + c +$ X and $\gamma + b +$ X production cross sections at $\sqrt{s} = 1.96$ TeV. An analysis for proton-proton and proton-lead collisions at the LHC regime is also performed.
\end{abstract}

\maketitle

\section{Introduction}
The inclusive direct (prompt) photon production in hadronic collisions has been quite useful for accessing information on parton distributions in hadrons. Moreover, direct photons appears to be an useful probe of the initial state of matter created in heavy ion collisions. Their interaction with the medium is purely electromagnetic, which gives a baseline for the interpretation of jet-quenching models. From the theoretical point of view, the prompt photon production is reasonably described in the next-to-leading order (NLO) perturbative QCD (pQCD) approach \cite{NLOQCD_pf} and on the color dipole formalism \cite{Boris,Boris2,Boris3,Boris4,Boris5,Mariotto} as well. Recently, the first measurements of inclusive photon production in association with heavy flavour jets have been reported by D0 Collaboration \cite{PRL}. Such a process provide important information on the parton content of the initial state hadrons as it is sensitive to charm (bottom) and gluon densities within the colliding hadrons.

In pQCD the inclusive production of prompt photons plus a heavy flavour quark \cite{Stavreva} is driven by the QCD Compton scattering, $g+Q\rightarrow \gamma +Q$, and also receives contribution from quark-antiquark annihilation process, $q+\bar{q}\rightarrow \gamma+g\rightarrow \gamma+Q\bar{Q}$. This is the reason of such a process to be sensitive to the heavy quark and gluon content of hadron, which have considerable uncertainties. The data description was found to be fairly good \cite{PRL} for large transverse momenta photons, where a perturbative approach is completely justifiable.   On the other hand, the same process can be addressed in the target rest frame \cite{Kop,BHQ}. In this framework, the direct photon production is viewed as an electromagnetic bremsstrahlung of a (light or heavy) quark that interacts with the target via gluonic exchanges.  This idea has been used to describe Tevatron data on prompt photon production at midrapidities using light quarks contribution \cite{Boris,Mariotto}.  Investigations of such formalism in association with aspects of saturation physics, for instance the Color Glass Condensate framework \cite{Raju},  have been made for $dA$ and $AA$ collisions in a number of works \cite{Jamal1,Jamal2,VB}.

In this work we apply the target rest frame formalism to photon plus heavy quark production cross section in high energies. The dependence of the production cross section on the final state particle rapidities and transverse momenta is investigated. In next section we introduce the main formulae to compute the hadronic differential cross section. It is already known from previous studies \cite{Mariotto} that the  photon transverse momentum spectrum is a sensitive probe of color dipole scattering cross section. We investigate the consequence of using several implementations of dipole cross sections which are constrained from deep inelastic scattering data. The theoretical results are compared to Tevatron  measurements  for the $\gamma + Q +$ X  production cross sections. An analysis for proton-proton and proton-lead collisions at the LHC regime is also developed. In order to do so, we will rely on geometric scaling arguments to do the transition from a nucleon to a nucleus target. The numerical results are discussed in detail in last section.

\section{Photon in association with heavy quarks in high energy hadron collisions}\label{sec:model}

In this section we summarize the relevant formulae to compute the $Q+\gamma$ production cross section at small-$x$ regime in the target rest frame. Such an approach has been early developed by Kopeliovich \cite{Kop} to the Drell-Yan production in the context of nuclear shadowing and further developed in \cite{BHQ}. Here, it will be considered that the photon transverse momentum is large compared to the hadronic scale, $\Lambda_{QCD}$, and at the same time smaller than the hadronic center-of-mass energy, $\Lambda_{QCD}^2 \ll {\pg}^2 \ll s$. In the target rest frame the production process is viewed as follows. A large-$x$ heavy quark (or heavy antiquark) of the hadron projectile scatters off the gluonic field of the hadron target and radiates a real photon. The relevant diagrams are those where the photon is radiated before or after the interaction with the target, whereas diagrams involving the quarks interacting with the target both before and after the photon vertex are suppressed in the high energy limit.  Although in the process of electromagnetic bremsstrahlung by a heavy-quark no quark-antiquark dipole participates, we soon observe that the cross section can be expressed via the  elementary dipole cross section, $N_{q\bar{q}}(x,r,b)$, of interaction of a $q\bar{q}$ dipole with a target. This allows us to make use of the extensive phenomenology on deep-inelastic scattering where the dipole cross section is very well determined from data on small-$x$ region \cite{Golec}.

The scattering cross section for production of a massless on-shell quark with momentum $\vec{l}$ and a real photon with momentum $\vec{k}$ was derived in  \cite{Jamal1,Jamal2}. The differential cross section for the process $ Q(p)+ h \rightarrow [ Q(l)\,\gamma(k)]+ X$ is given by:
\begin{eqnarray}
& & \frac{d^3\hat{\sigma}}{d(\pg)^2\, dy_{\gamma}\, dy_Q}  =  \int \int d(\pq)^2\, d\Delta \phi\,H\left(\vec{p_{\perp}}^{\gamma},\vec{p_{\perp}}^{Q} ,y_{\gamma},y_Q\right) \nonumber \\
& & \times \, {\cal N}(x_g,\vec{p_{\perp}}^{\gamma},\vec{p_{\perp}}^{Q})\,\delta \left(x_Q - \frac{\pg e^{y_{\gamma}}}{\sqrt{s}}-\frac{\pq e^{y_{Q}}}{\sqrt{s}} \right),
\label{eq:cs_gen}
\end{eqnarray}
with the quantity $H$ being defined as,
\begin{eqnarray}
H\left(\vec{p_{\perp}}^{\gamma},\vec{p_{\perp}}^{Q} ,y_{\gamma},y_Q\right)  & = &  \frac{e_q^2\, \alpha_{em}}{\sqrt{2}\,(2\pi)^3} \,
\left[1 + \left(\frac{\beta_Q}{\beta_Q+\beta_{\gamma}}\right)^2\right]\nonumber \\ 
 & \times & \,\frac{\beta_{\gamma}^2}{(\pg)^2 \sqrt{s}} \frac{\left(\vec{p_{\perp}}^{\gamma}+\vec{p_{\perp}}^Q \right)^2}{\left(\frac{\beta_{\gamma}}{\beta_Q}\, \vec{p_{\perp}}^Q - \vec{p_{\perp}}^{\gamma}\right)^2},\nonumber \\
\label{eq:cs_gen2}
\end{eqnarray}
where the incoming heavy quark has momentum $p$, the photon and outgoing quark rapidities are defined via $\beta_{\gamma}\equiv k^- =\frac{\pg}{\sqrt{2}}\, e^{y_{\gamma}}$ and $\beta_Q\equiv l^- = \frac{\pq}{\sqrt{2}}\, e^{y_Q}$. The corresponding transverse momenta of photon and outgoing quark are $\pg$ and $\pq$, respectively. The angle $\Delta \phi$ is the opening angle between the final state quark and photon defined as $\cos \,(\Delta \phi ) \equiv (\vec{p_{\perp}}^{\gamma} \cdot \vec{p_{\perp}}^{Q})/\pg\,\pq$, with respect to the produced quark axis. The quantity ${\cal N}$ is related to the color dipole cross section, $N(x,r)$, which should satisfy the QCD evolution equations and that includes the small-$x$ evolution. It is defined as:
\begin{eqnarray}
\label{eq:cs_def}
{\cal N}(x_g,\vec{p_{\perp}}^{\gamma},\vec{p_{\perp}}^{Q}) & = & \int d^2b\int d^2r\, e^{i\vec{r}\cdot \vec{p_T}}N(x_g,r,b),\\
& = & (\pi R_h^2)\,\int 2\pi rdr J_0(r\,p_T)\,N(x_g,r),\nonumber
\end{eqnarray}
where the $b$ is the impact paramater of interaction and $r$ is the corresponding transverse size. Here, we have used the notation $\vec{p_T}=(\vec{p_{\perp}}^{\gamma}+\vec{p_{\perp}}^Q)$ and $p_T=|\vec{p_T}|$. In last line in Eq. (\ref{eq:cs_def}) we treat the target (proton)  as a homogeneous disk of radius $R_h\simeq 5$ GeV. The momentum fraction  $x_g$ is related to the photon and final state quark rapidities and transverse momenta as follows \cite{Jamal2}:
\begin{eqnarray}
x_g = \frac{\pg}{\sqrt{s}}\,e^{-y_{\gamma}} + \frac{\pq}{\sqrt{s}}\,e^{-y_Q}.
\label{eq:x_g}
\end{eqnarray}

In order to compute the hadronic cross section for photon plus heavy quark production it is needed to convolute the partonic cross section, Eq. (\ref{eq:cs_gen}),  with the heavy quark  distribution function (PDF) on a proton, $f_Q(x,Q)$. Therefore, the differential cross section for the hadronic production is given by,
\begin{eqnarray}
\frac{d^3\sigma \,(pp\rightarrow \gamma Q+X)}{d(\pg)^2\, dy_{\gamma}\, dy_Q} & = &\int dx_Q\,f_Q(x_Q,\mu^2)\, \frac{d^3\hat{\sigma}}{d(\pg)^2\, dy_{\gamma}\, dy_Q},\nonumber\\
& & 
\label{eq:final}
\end{eqnarray}

The charm and bottom PDFs are presently assumed to be radiatively generated  and then are related to the gluon PDF through DGLAP evolution equations. It is noticed that the charm structure function at large-$x$ measured by EMC Collaboration suggests that there might be an intrinsic charm (IC) component  in nucleon. In general, nonperturbative models are considered to describe such a component (see, for instance Refs. \cite{ICs,IC2,IC3}). Those models are not considered here. In Eq. (\ref{eq:final}), the difference between charm and bottom comes from the quark charge, $e_c^2=4/9$ and $e_b^2=1/9$, and from the quark PDF (charm PDF is larger than the bottom one). For the factorization/renormalization scale, $Q=\mu$, the photon transverse momentum  is usually chosen.

To proceed further, we need to know the quantity ${\cal N}(x_g,\vec{p_{\perp}}^{\gamma},\vec{p_{\perp}}^{Q})$. It can be computed numerically using the available phenomenological dipole cross sections \cite{Golec}. Here, we consider some of them that have been used to describe deep inelastic scattering (DIS) data and RHIC data as well. We can gain some physical insight using analytical versions of the amplitude ${\cal N}$. For instance, considering small dipole configurations (which is the typical configuration for large $p_T$ considered in the current study) the color dipole amplitude behaves like $N(r\rightarrow 0)\approx (r^2Q_s)^{\gamma}/4$.  Such an expression is typical in saturation models based on the QCD nonlinear  evolution equations for the dipole amplitude and its behavior is known as geometric scaling on the variable $rQ_s$. The intrinsic momentum scale, $Q_s\propto x_g^{-\lambda}$,  is the so-called saturation scale (For central rapidities at Tevatron, $Q_s\leq 1$ GeV). For some fixed values of the anomalous dimension, $\gamma$, the zeroth-order Hankel transform in Eq. (\ref{eq:cs_def}) can be done analytically considering $p_T>0$. In Ref. \cite{Jamal2}, it has been shown that for $\gamma =1/2$ (BFKL anomalous dimension) the result is the following:
\begin{eqnarray}
 \frac{{\cal N}(x_g,\vec{p_{\perp}}^{\gamma},\vec{p_{\perp}}^{Q})}{\pi\,R_h^2} =
 \frac{32\,\pi }{Q_s^2}\, \left[1 + \frac{16 \left|\vec{p_{\perp}}^{\gamma} + \vec{p_{\perp}}^Q\right|^2}{Q_s^2}\right]^{-3/2}.
 \label{eq:N_approx}
 \end{eqnarray}
The current Tevatron data on $\gamma + Q$ production is dominated by large photon transverse momentum, $\pg \geq 30$ GeV $\gg Q_s$, and then it is theoretically expected that the anomalous dimension is close to the DGLAP values, $\gamma_{\mathrm{DGLAP}}=1$. Thus, it is important to consider dipole models where the anomalous dimension is running. This is named extended geometric scaling property. For a running anomalous dimension which does not depend explicitly on the dipole size $r$, it can be obtained the following \cite{BUW}: 
\begin{eqnarray}
 \frac{{\cal N}(x_g,p_T)}{\pi\,R_h^2} \approx
 \frac{\pi\,\left(2\,Q_s\right)^{2\gamma(\omega)}}{p_T^{2\gamma(\omega)+2}}\, \frac{\Gamma \left(1+\gamma(\omega)\right)}{-\Gamma \left(-\gamma(\omega)\right)},
 \label{eq:N_wessels}
 \end{eqnarray}
where we consider two models for the running anomalous dimension, which have been confronted to RHIC data on charged particle multiplicities. The first one is the BUW model \cite{BUW}, whereas the second one is the DHJ model \cite{DHJ}. The expression for $\gamma (\omega)$, with $\omega=p_T/Q_s$, for each model is given by:
\begin{eqnarray}
\gamma_{\mathrm{BUW}} & = & \gamma_s + (1-\gamma_s)\,\frac{(\omega^a-1)}{(\omega^a-1)+b}, \\
\gamma_{\mathrm{DHJ}} & = & \gamma_s + (1-\gamma_s)\,\frac{|\ln\omega^2|}{|\ln \omega^2|+\lambda Y+d\sqrt{Y}}.
\end{eqnarray}
The BUW parameters are $a=2.82$ and $b=168$ \cite{BUW} and the DHJ one is $d=1.2$ (in addition, one has $Y=\ln(1/x)$)  \cite{DHJ}. In both cases, $\gamma_s=0.628$, $\lambda=0.3$ and $x_0=3\cdot 10^{-3}$, with a saturation scale on the nucleon given by $Q_s^2= (x_0/x_g)^{\lambda}$ GeV$^2$.

Furthermore, in Ref. \cite{VM} it has been shown that the quantity ${\cal N}$ is directly connected to the unintegrated gluon distribution, ${\cal F}(x,k_{\perp})$, in the small dipole size ($r\rightarrow 0$) limit.
 Using such a connection,  we can now use the available numerical/analytical implementations of the unintegrated gluon function. In next section, we calculate the differential cross section, Eq. (\ref{eq:final}), using the analytical expressions for the dipole amplitude, Eq. (\ref{eq:N_wessels}),  computed from BUW and DHJ  models and a numerical result for the unintegrated gluon distribution. We will discuss the main features coming from distinct theoretical approaches when compared to the NLO perturbative QCD calculations. A study for an extrapolation to $pp$ and $pA$ collisions ate the LHC will be also presented.

\begin{figure}[t]
\includegraphics[scale=0.45]{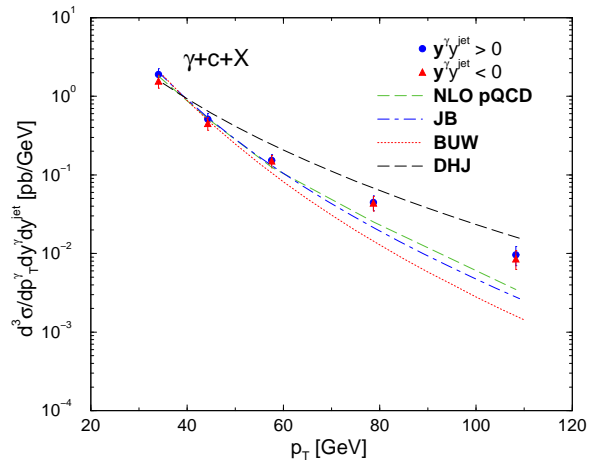}
\caption{The $\gamma+c+X$ differential cross section as a function of $p_{\perp}^{\gamma}$. Experimental data from D0 Collaboration \cite{PRL} (only systematic errors are presented). }\label{fig:1}
\end{figure}

\begin{figure}[t]
\includegraphics[scale=0.45]{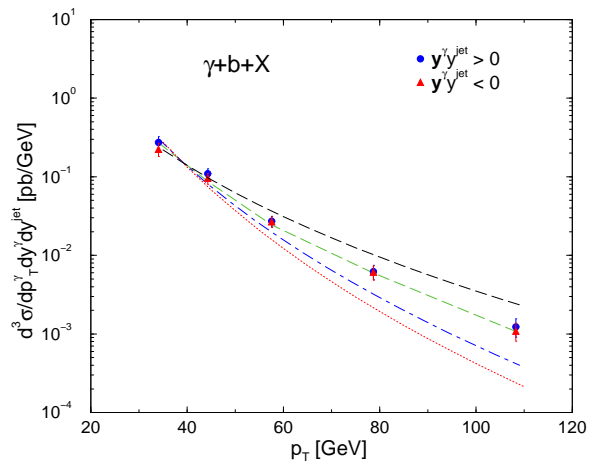}
\caption{The $\gamma+b+X$ differential cross section as a function of $p_{\perp}^{\gamma}$. Experimental data from D0 Collaboration \cite{PRL} (only systematic errors are presented). }\label{fig:2}
\end{figure}

\section{Discussions and summary}\label{sec:ccl}
Let us now compare the theoretical calculations presented in previous section  to the experimental data for  production cross sections in $p\bar{p}$ collisions at Tevatron for center of mass energy $\sqrt{s}=1.96$ TeV. Recently, D0 Collaboration have performed the first measurement of the differential cross section of inclusive photon production in association with heavy flavors jets \cite{PRL}. The results cover the range in photon transverse momentum $30<p_{\perp}^{\gamma}<150$ GeV ad  photon/jet rapidities, $|y_{\gamma}|<1$ and $|y_{\mathrm{jet}}|<0.8$. The following cut is required to jet transverse momentum $p_{\perp}^{\mathrm{jet}}> 15$ GeV. For simplicity, here we will take central rapidities $y_{\gamma},\,y_{Q}=0$ and consider, $y_{\mathrm{jet}}=y_Q$. Concerning the numerical calculation of differential cross sections in  Eq. (\ref{eq:cs_gen}) some comments are in order. As already discussed in Ref. \cite{Jamal2}, despite any particular assumptions on the dipole amplitude the factor in denominator of Eq. (\ref{eq:cs_gen2}) diverges as the momenta of produced quark and photon are parallel, $\Delta\phi= 0$, which is the usual collinear divergence present in perturbation theory. Thus, in our calculations we consider a lower cut-off for the angular integration, $\Delta \phi_{\mathrm{min}}$. Moreover, a numerical calculation of the Hankel transform of zero order in Eq. (\ref{eq:cs_def}) is divergent when  $|\vec{p_{\perp}}^{\gamma} + \vec{p_{\perp}}^{Q}| = 0$, which occurs for $\Delta \phi = \pi$ if momenta are equal, $\pg=\pq$. In our case, this is not a shortcoming as analytical expressions for the color dipole amplitudes are  being used. For a similar discussion connected to the present study on the angular dependence of production cross section we quote the recent work in Ref. \cite{Jamal3}. As a last comment, we are using the target rest frame approach in the limit of its validity in the Tevatron regime. Formally, this approach is valid at small-$x_g$, see Eq. (\ref{eq:x_g}), and a careful analysis on what is the $x_g$-range being probed is difficult as the  quark transverse momentum, $\pq$, is integrated over in  Eq. (\ref{eq:cs_gen}). However, the minimum $x_g$ value can be estimated. Using the Tevatron quark transverse momentum cut, $\pq >15$ GeV and the photon transverse momentum range, $30<\pg <150$ GeV one gets $0.02 \leq x_g^{\mathrm{min}}\leq 0.08$ at central rapidities.

In Figs. \ref{fig:1} and \ref{fig:2} the $\gamma+c+X$ and  $\gamma+b+X$ differential cross section are presented as a function of $p_{\perp}^{\gamma}$. Experimental data from D0 Collaboration \cite{PRL} are shown, where only the systematic errors are presented. The long-dashed lines (in green) represent the the NLO pQCD predictions \cite{Stavreva}  using CTEQ6.6M PDFs \cite{CTEQ} (averaged over rapidity regions), where the renormalization scale $\mu_{R}$, factorization scale $\mu_{F}$ and fragmentation scale $\mu_f$ are all set equal to $\pg$. In our numerical calculations the same set is used, i.e. in Eq. (\ref{eq:final}) we consider $\mu^2=(\pg)^2$. The long-dashed curves (in blue) correspond to the calculation using the connection of the dipole amplitude ${\cal N}$ to the unintegrated gluon function. For the unintegrated gluon PDF we consider the J. Bluemlein parametrization (JB) \cite{JB}, which relies on BFKL approach \cite{bfkl}:
\begin{eqnarray}
 \label{conv}
 {\cal F}(x_g,p_T^2,\mu^2) = \int_{x_{g}}^1
 {\cal G}(\eta,p_T^2,\mu^2)\,
 \frac{x_g}{\eta}\,g(\frac{x_g}{\eta},\mu^2)\,d\eta,
\end{eqnarray}
where it is defined the ${\cal G}$ function as,
\begin{eqnarray} \label{J0}
 {\cal G}(\eta,p_T^2)=\frac{\bar{\alpha}_s}{\eta\,p_T^2}\,
 J_0\left(2\sqrt{\bar{\alpha}_s\ln(1/\eta)\ln(\mu^2/p_T^2)}\right),
 \, (p_T^2<\mu^2) \nonumber \\
 \label{I0}
 {\cal G}(\eta,p_T^2)=\frac{\bar{\alpha}_s}{\eta\,p_T^2}\,
 I_0\left(2\sqrt{\bar{\alpha}_s\ln(1/\eta)\ln(p_T^2/\mu^2)}\right),
 \, (p_T^2>\mu^2), \nonumber
\end{eqnarray}
where $J_0$ and $I_0$ stand for Bessel functions (of real and imaginary
arguments, respectively), and $\bar{\alpha}_s=\alpha_s/3\pi$. The LO MRST
set \cite{MRST} was used in our calculations as the input gluon collinear
density, $g(x,Q)$.

Let us now comment on the numerical results for theoretical approaches considered here. As already found in Ref. \cite{PRL}, the NLO pQCD calculation describes in good agreement the bottom data. However, the large $\pg$ range of charm data is underestimated and it seems to favor a IC component which would be important at that region. On the other hand, the target rest frame approach produces distinct results depending on the input for the dipole cross section. The JB parameterization is close to the NLO pQCD (the anomalous dimension for JB is near the DGLAP one as it is derived in the double logarithmic, DLL, limit of BFKL approach) for charm and deviates from pQCD at $\pg \geq $ 60 GeV for the bottom case. A weak reason could be due the fact we are considering the strong coupling fixed, $\alpha_s=0.2$, in JB parameterization for both cases and a distinct bottom PDF. We have checked that it is not the case. Probably, the deviation does come from the relative contribution of a set of diagrams that are not accounted for in the target rest frame approach compared to pQCQ. For instance,  it is shown in Ref. \cite{Stavreva} that the annihilation process $q\bar{q}\rightarrow Q\bar{Q}\gamma$ is very important at large $\pg$ for the bottom case. The target rest frame approach only involves the Compton subprocess and can not be directly compared to the NLO pQCD calculation.

The BUW and DHJ parameterizations differ not only in the scaling behavior, but in the way the large $p_T$ limit of $\gamma(\omega)$ approaches to one. This last feature leads to different large momentum slopes of dipole amplitude and therefore to distinct predictions for the large $\pg$-slope when using Eqs. (\ref{eq:final}), (\ref{eq:cs_gen2}) and (\ref{eq:N_wessels}). For instance, in \cite{BUW} it was shown that for BUW model, ${\cal N}_{\mathrm{BUW}}(\gamma\rightarrow 1)\propto (Q_s/p_T)^{2+a}/p_T^2$, whereas for DHJ one has  ${\cal N}_{\mathrm{DHJ}}(\gamma\rightarrow 1)\propto Q_s^{2+a}/[p_T^4\ln(p_T^2/Q_S^2)]$. Therefore, the future data on forward production (where current approach is formally valid even at large $\pg$)  of heavy quarks in association with prompt photon could discriminate among the distinct models for the dipole cross section.

\begin{figure}[t]
\includegraphics[scale=0.45]{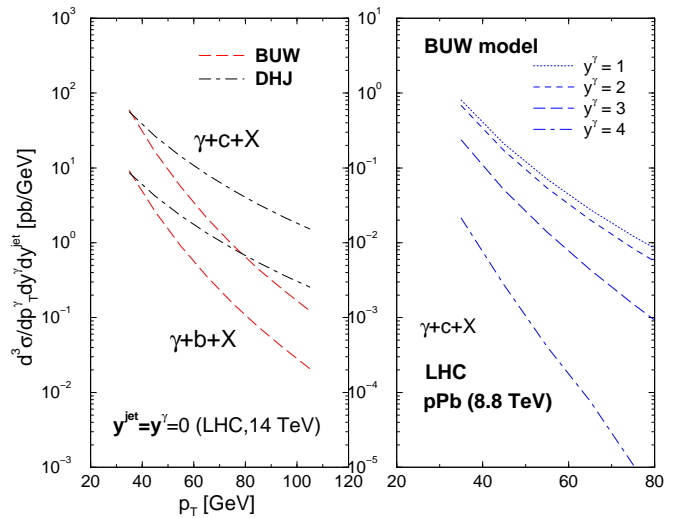}
\caption{The $\gamma+Q+X$ differential cross section at LHC regime. Left panel: proton-proton  cross sections at central jet/photon rapidities as a function of $p_{\perp}^{\gamma}$ at $\sqrt{s}=14$ TeV. Right panel: proton-lead (pPb) cross section for charm plus photon for fixed photon rapidities $y^{\gamma}=1,2,3,4$ at $\sqrt{s}=8.8$ TeV.}\label{fig:3}
\end{figure}

Finally, we perform some estimates for the LHC kinematic regime. Let us start by $pp$ collisions at energy of $\sqrt{s}=14$ TeV. Now, even at central rapidities the $x_g$ is sufficiently small to justify the target rest frame approach at large transverse momenta. Keeping the same kinematic cuts from Tevatron, $\pq >15$ GeV and  $30<\pg <150$ GeV, one gets $0.003 \leq x_g^{\mathrm{min}}\leq 0.01$. The situation is further improved in the forward rapidities case. In Fig. \ref{fig:3} (left panel) are shown the estimates for the differential cross section (for charm and bottom) as a function of photon transverse momentum at central rapidities. The dashed curves represents the results for BUW model, whereas the dot-dashed ones stand for the DHJ model. The deviations at large $\pg$ follow the same pattern as for Tevatron. The conclusions are similar as previously discussed, where the transverse momentum distribution allows discrimination among phenomenological/theoretical models for the dipole scattering cross section (or for the unintegrated gluon distribution in an indirect way). At the LHC the center-of-mass energy is sufficiently high and annihilation process no longer dominates at large $\pg$ as seen for Tevatron, mostly for the bottom case.

In Fig. \ref{fig:3} (right panel) we present the proton-lead (pPb) differential cross section at $\sqrt{s}=8.8$ TeV as a function of photon transverse momentum for fixed forward photon rapidities $y^{\gamma}=1,2,3,4$ (from the top to bottom, respectively). In order to obtain the cross section for a nuclear target we rely on the geometric scaling arguments \cite{GS}: it is replaced $R_p \rightarrow R_A$ in Eq. (\ref{eq:cs_def}) and also $Q_{s,p}^2\rightarrow (AR_p^2/R_A^2)^{\Delta}\,Q_{s,p}^2$, where $\Delta =1.26$. In case of $\Delta=1$ that replacement becomes the usual assumption for the nuclear saturation scale, $Q_{s,A}^2=A^{1/3}\,Q_{s,p}^2$. Such an approach has been used in Ref. \cite{BM} to describe small-$x$ data on the nuclear structure functions.  This enhances the saturation scale by a factor six for a lead nucleus. As photon rapidity diminishes, i.e. smaller $x_g$, the saturation scale increases since $Q_s\propto x_g^{-\lambda}$. Such a behavior implies an enhancement on the quantity $\omega=p_T/Q_s$ at fixed $\pg$ in Eq. (\ref{eq:N_wessels}) and therefore large rapidities means bigger $\omega$ and the running anomalous dimension increases towards $\gamma\rightarrow 1$. This patters is viewed in the plot corresponding to a steep $\pg$-slope as rapidities increase. We have checked that the $\pg$-slope is unchanged by considering forward quark rapidities, $y_{\mathrm{jet}}=y_Q\geq 1$, at fixed central photon rapidity. As a final comment, the $\gamma + Q$ production in pPb is important {\it per se} as the nuclear gluon distribution is largely unconstrained and such a process can test at the same time the gluon and charm/bottom distribution functions. Thus, measurements having appropriated experimental  error could distinguish the distinct parameterizations for the nuclear PDFs.

As a summary, we have applied the target rest frame formalism to photon plus heavy quark production cross section in hadronic collisions at high energies. The dependence of the production cross section on the photon and quark rapidities and transverse momenta have been investigated. It was verified that the photon transverse momentum slope is a good probe of color dipole scattering amplitude as it directly depends on the (running) anomalous dimension, which is determined from the underlying QCD dynamics in a given kinematic regime. The theoretical results are first compared to Tevatron  measurements of the differential $\gamma + Q +$ X production cross sections having in mind that this is an extrapolation on kinemetic regime where the small-$x_g$ limit id not completely fulfilled. An analysis for proton-proton and proton-lead collisions at the LHC regime, including the forward rapidity region, has been performed. For the $pA$ case, a model for the saturation scale for a nucleus has been introduced based on geometric scaling  arguments, where $Q_s^2(x_g,A)\propto A^{1/3}Q_s^2(x_g)$. This procedure can be also done using a nuclear version of the unintegrated gluon distribution as proposed for instance in \cite{BM}.

\section*{Acknowledgments}
This work was supported by the funding agencies CNPq and FAPERGS, Brazil. One of us (MVTM) thanks Tzvetalina Stavreva for discussion and helpful comments.

\end{document}